\documentstyle{mn}
\input psfig.tex

\newif\ifAMStwofonts

\ifoldfss
  \ifCUPmtlplainloaded \else
    \NewTextAlphabet{textbfit} {cmbxti10} {}
    \NewTextAlphabet{textbfss} {cmssbx10} {}
    \NewMathAlphabet{mathbfit} {cmbxti10} {} % for math mode
    \NewMathAlphabet{mathbfss} {cmssbx10} {} %  "   "    "
  \fi
  \ifAMStwofonts
    \ifCUPmtlplainloaded \else
      \NewSymbolFont{upmath} {eurm10}
      \NewSymbolFont{AMSa} {msam10}
      \NewMathSymbol{\upi}     {0}{upmath}{19}
      \NewMathSymbol{\umu}     {0}{upmath}{16}
      \NewMathSymbol{\upartial}{0}{upmath}{40}
      \NewMathSymbol{\leqslant}{3}{AMSa}{36}
      \NewMathSymbol{\geqslant}{3}{AMSa}{3E}

    \fi
  \fi
\fi % End of OFSS

\ifnfssone
  \newmathalphabet{\mathit}
  \addtoversion{normal}{\mathit}{cmr}{m}{it}
  \addtoversion{bold}{\mathit}{cmr}{bx}{it}
  \newmathalphabet{\mathbfit} % math mode version of \textbfit{..}
  \addtoversion{normal}{\mathbfit}{cmr}{bx}{it}
  \addtoversion{bold}{\mathbfit}{cmr}{bx}{it}
  \newmathalphabet{\mathbfss} % math mode version of \textbfss{..}
  \addtoversion{normal}{\mathbfss}{cmss}{bx}{n}
  \addtoversion{bold}{\mathbfss}{cmss}{bx}{n}
  \ifAMStwofonts
    \ifCUPmtlplainloaded \else
      \UseAMStwoboldmath
      \makeatletter
      \new@mathgroup\upmath@group
      \define@mathgroup\mv@normal\upmath@group{eur}{m}{n}
      \define@mathgroup\mv@bold\upmath@group{eur}{b}{n}
      \edef\UPM{\hexnumber\upmath@group}
      \new@mathgroup\amsa@group
      \define@mathgroup\mv@normal\amsa@group{msa}{m}{n}
      \define@mathgroup\mv@bold\amsa@group{msa}{m}{n}
      \edef\AMSa{\hexnumber\amsa@group}
      \makeatother
      \mathchardef\upi="0\UPM19
      \mathchardef\umu="0\UPM16
      \mathchardef\upartial="0\UPM40
      \mathchardef\leqslant="3\AMSa36
      \mathchardef\geqslant="3\AMSa3E
    \fi
  \fi
\fi % End of NFSS release 1

\ifnfsstwo
  \DeclareMathAlphabet{\mathbfit}{OT1}{cmr}{bx}{it}
  \SetMathAlphabet\mathbfit{bold}{OT1}{cmr}{bx}{it}
  \DeclareMathAlphabet{\mathbfss}{OT1}{cmss}{bx}{n}
  \SetMathAlphabet\mathbfss{bold}{OT1}{cmss}{bx}{n}
  \ifAMStwofonts
    \ifCUPmtlplainloaded \else
      \DeclareSymbolFont{UPM}{U}{eur}{m}{n}
      \SetSymbolFont{UPM}{bold}{U}{eur}{b}{n}
      \DeclareSymbolFont{AMSa}{U}{msa}{m}{n}
      \DeclareMathSymbol{\upi}{0}{UPM}{"19}
      \DeclareMathSymbol{\umu}{0}{UPM}{"16}
      \DeclareMathSymbol{\upartial}{0}{UPM}{"40}
      \DeclareMathSymbol{\leqslant}{3}{AMSa}{"36}
      \DeclareMathSymbol{\geqslant}{3}{AMSa}{"3E}
    \fi
  \fi
\fi % End of NFSS release 2

\ifCUPmtlplainloaded \else
  \ifAMStwofonts \else % If no AMS fonts
    \def\upi{\pi}
    \def\umu{\mu}
    \def\upartial{\partial}
  \fi
\fi

\def\simlt{\lower.5ex\hbox{$\; \buildrel < \over \sim \;$}}
\def\simgt{\lower.5ex\hbox{$\; \buildrel > \over \sim \;$}}
\def\ms{M$_{\odot}$}

\def\mp{M$_{\odot}$ pc$^{-2}$}

\begin{document}

\title{Chemo-spectrophotometric evolution of spiral galaxies: \\
     III. Abundance and colour  gradients in discs}

\author[N. Prantzos and S. Boissier]
       {N. Prantzos and S. Boissier \\
 Institut d'Astrophysique de Paris, 98bis, Bd. Arago, 75104 Paris}
\date{ }

\pagerange{\pageref{firstpage}--\pageref{lastpage}}
\pubyear{1999}
\maketitle

\label{firstpage}

\begin{abstract}

We study the relations between luminosity and chemical abundance profiles
of spiral galaxies, using detailed models for the chemical and
spectro-photometric evolution of galactic discs. The models are ``calibrated''
on the Milky Way disc and are successfully 
extended to other discs with the help of
simple ``scaling'' relations, obtained in the framework of semi-analytic models
of galaxy formation.
We find that our models exhibit oxygen abundance gradients that increase in absolute value
with decreasing disc luminosity (when expressed in dex/kpc) and are independent of
disc luminosity (when expressed in dex/scalelength), both in agreement with
observations. We notice an important  strong correlation between abundance gradient 
and disc scalelength. These results support the idea of ``homologuous evolution''
of galactic discs.   
\end{abstract}

\begin{keywords}
Galaxies: general - evolution - spirals - photometry - abundances 
\end{keywords}

\section{Introduction}

Almost all large spiral galaxies present sizeable radial abundance
gradients. This is, for instance, the case for the Milky Way disc, showing an 
oxygen abundance gradient of dlog(O/H)/dR$\sim$-0.07 dex/kpc in both its 
gaseous and stellar components. The origin of these gradients is still a matter
of debate. A radial variation of the star formation rate (SFR), or 
the existence of radial gas flows, or a combination of these processes,
can lead to abundance gradients in discs
(e.g. Lacey and Fall 1985, Koeppen 1994, Edmunds and Greenhow 1995,
 Tsujimoto et al. 1995, Firmani et al. 1996, Molla et al.  1997,
Matteucci and Chiappini 1998). On the other hand, the presence
of a central bar inducing large scale mixing through radial gas flows
tends to level out preexisting abundance gradients
(e.g. Dutil and Roy 1999). Despite several studies
in the 90ies (e.g. Friedli et al. 1998 and references therein) 
the relative importance of these processes has not been clarified yet.

Several large spectophotometric surveys and literature compilations have 
appeared in recent years (Vila-Costas and Edmunds 1992, Zaritsky et al. 1994,
Garnett et al. 1997, van Zee et al. 1998). These studies allowed to establish several
important features of abundances          in spirals (Henry and Worthey 
1999, and references therein). In particular, it is established now that
the absolute abundance at a given galactocentric distance is correlated
to galaxy luminosity (and rotational velocity). Also, when abundance gradients
are expressed in dex/kpc there is a tendency of smaller discs to exhibit
larger gradients; but when they are expressed in dex/R$_d$ (where R$_d$ is
the disc exponential scalelength), that correlation disappears (Garnett et 
al. 1997).

In a series  of recent papers (Boissier and Prantzos 1999a, 1999b; hereafter
BP99a and BP99b, respectively) we presented a detailed model
of the chemical and spectrophotometric evolution of spiral galaxies.
Adopting some simple prescriptions for the star formation and infall rates,  
we first applied the model to the Milky Way disc and showed that it 
reproduces all major observables of our Galaxy (BP99a), including
radial profiles of gas, stars, SFR and oxygen abundances.  
Assuming that the Milky Way is a typical spiral galaxy, we
extended then the model to the study of other spirals (BP99b)
with the help of some simple ``scaling laws'', which have been established 
in the framework of the Cold Dark Matter scenarios for galaxy formation.
We showed  that this simple ``scaled'' model reproduces quite
satisfactorily most of the observed integrated properties of 
present day spirals:
disc sizes and central surface brigthness, Tully-Fisher relations in various 
wavelength bands, colour-colour and colour-magnitude relations, gas fractions 
vs. magnitudes and colours, abundances vs. local and integrated properties,
as well as spectra for different galactic rotational velocities. 
Despite the extremely simple nature of our models, we find satisfactory
agreement with all those observables, provided the timescale for star
formation in low mass discs is longer than for more massive ones.

In this third paper of the series we explore the implications of our models
for the radial profiles of spiral galaxies and, in particular, the
abundance and photometry gradients.
The plan of the paper is as follows: In Sec. 2 we present the basic 
ingredients and the underlying assumptions of the model. 
The results of the models concerning the evolution of the radial oxygen 
abundance profiles and the final abundance and luminosity gradients are
presented in Sec. 3. In Sec. 4 we compare our results to observations;
in particular, we show that there is a clear correlation between abundance
gradient expressed in dex/kpc and disc scalelength, both observationnaly
and in our models. The implications of these results for disc evolution are
discussed in Sec. 5.

\section{The model}

In this section we recall the main ingredients and the underlying asumptions
of our chemical and spectrophotometric evolution  models.
 
 The galactic disc is simulated as an ensemble of concentric, independently
evolving rings, gradually built up by infall of primordial composition. The
chemical evolution of each zone is followed by solving the appropriate
set of integro-differential equations (Tinsley 1980), 
without the Instantaneous Recycling
Approximation. Stellar yields are from Woosley and Weaver (1995) 
for massive stars
and Renzini and Voli (1981) for intermediate mass stars. Fe producing SNIa are
included, their rate being calculated with the prescription of Matteucci and
Greggio (1986). The adopted stellar Initial Mass Function (IMF)
is a multi-slope power-law between 0.1 \ms \ and 100 \ms \ from the work of
Kroupa et al. (1993).

The spectrophotometric evolution is followed in a self-consistent way, i.e.
with the  SFR $\Psi(t)$ and metallicity $Z(t)$ of each zone determined 
by the chemical evolution,
and the same IMF. The stellar lifetimes, evolutionary tracks and spectra are
metallicity dependent; the first two are from the Geneva library 
(Schaller  et al. 1992, Charbonnel et al. 1996) and the latter from 
Lejeune et al. (1997). Dust absorption is
included according to the prescriptions of  
Guiderdoni et al. (1998) and assuming a ``sandwich''
configuration for the stars and dust layers (Calzetti et al. 1994).

The star formation rate (SFR) is locally given by a
Schmidt-type law, i.e proportional to some power of the gas surface
density $\Sigma_g$ and varies with galactocentric radius $R$ as:
\begin{equation}
 \Psi(t,R) \ = \  \alpha \  \Sigma_g(t,R)^{1.5} \ V(R) \ R^{-1}
\end{equation}
where $V(R)$ is the circular velocity at radius $R$. This radial dependence of
the SFR is suggested by
the theory of star formation induced by density waves in spiral
galaxies (e.g. Wyse and Silk 1989).

The infall rate is assumed to be exponentially decreasing in time, i.e.
\begin{equation}
f(t,R) \  = \ A(R) \ e^{-t/\tau(R)}
\end{equation}
with $\tau(R_0$=8 kpc) = 7 Gyr in the solar neighborhood, 
in order to reproduce the local G-dwarf
metallicity distribution.
The coefficient $A(R)$ is obtained by the requirement that at time 
T=13.5 Gyr  the current
mass profile of the disc $\Sigma(R)$ is obtained, i.e.
\begin{equation}
\int_0^T f(t,R) \  = \ \Sigma(R)
\end{equation}
with 
\begin{equation}
 \Sigma(R) \ = \  \Sigma_0 \  e^{-R/R_d}  
\end{equation}
In the case of the Milky Way disc one has for the central surface density
$\Sigma_0$=1150 \mp \ and for the disc scalelength $R_d$=2.6 kpc,
leading to an overall disc mass $M_{d}\sim$5 10$^{10}$ \ms \ (see BP99a and
references therein).

The efficiency $\alpha$ of the SFR  (Eq. 1) is fixed
by the requirement that the observed local gas fraction
$\sigma_g(R_0$=8 kpc)$\sim$0.2,
is reproduced at T=13.5 Gyr. 
We consider then that the really ``free'' parameters
of the model are the radial dependence of the
infall timescale $\tau(R)$ and of the SFR.
It turns out that
the number of observables explained by the model is much
larger than the number of free parameters. In particular		
the model reproduces present day ``global'' properties 
(profiles of gas, O/H, SFR, and supernova rates), as well as	
the current disc luminosities in various wavelength bands 
and the corresponding radial profiles of gas, stars, SFR and metal abundances;
moreover, the adopted inside-out star forming scheme leads to a 
scalelength of $\sim$4 kpc in the B-band and $\sim$2.6 kpc in the K-band, 
in agreement with observations (see BP99a).

\begin{figure*}
\psfig{file=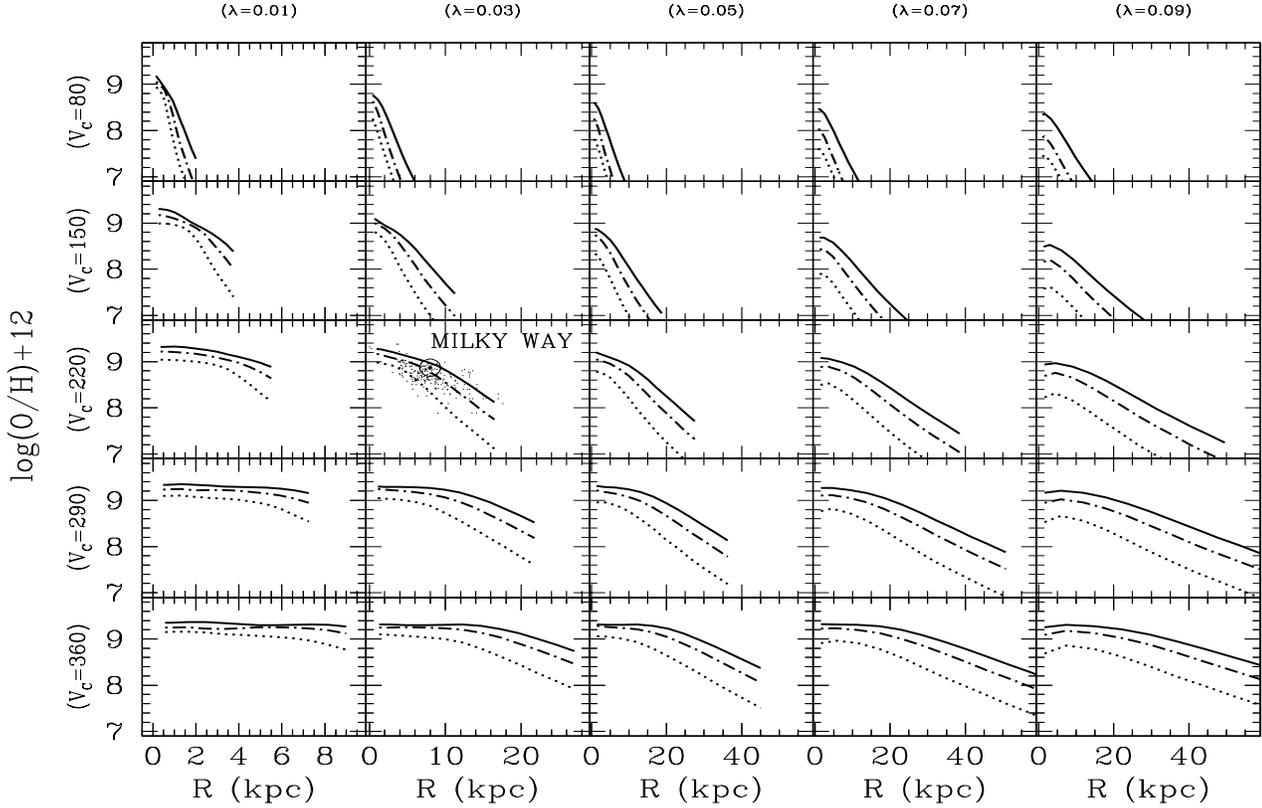,height=11.cm,width=\textwidth,angle=-90}
\caption{\label {}
Evolution of oxygen abundance profiles in our models, at times
t=3 Gyr ({\it dotted} curves), 7.5 Gyr ({\it dash-dotted} curves) and 13.5 Gyr 
({\it solid} curves).  Models are arranged according to values of maximum rotational
velocity $V_C$ (increasing from top to bottom) and spin parameter $\lambda$ (increasing
from left to right). In the case of the Milky Way model ($V_C$=220 km/s, $\lambda$=0.03)
observations of oxygen abundances in HII regions, planetary nebulae and B-stars are also
plotted; the observed metallicity gradient dlog(O/H)/dR$\sim$-0.07 dex/kpc is well 
reproduced. Notice that the absolute values of $\lambda$ do not enter directly the
calculations, since the model discs are {\it scaled} to the one of the Milky Way through the
relations (5) and (6); in other terms, if the Milky Way $\lambda$ is taken to be
0.06, the same results would be obtained for discs with $\lambda$ = 0.02, 0.06, 0.10, 0.14
and 0.18, respectively.}

\end{figure*}

In order to extend the model to other disc galaxies we adopt  the 
``scaling properties'' derived by Mo, Mao and White (1998, hereafter MMW98) 
in the framework of the Cold Dark Matter (CDM) scenario for galaxy formation. 
According to this scenario, primordial density fluctuations give rise to 
haloes of non-baryonic dark
matter of mass $M$, within which baryonic gas condenses later and forms discs
of maximum circular velocity $V_C$. 
It turns out that disc profiles can be expressed in terms of only two 
parameters: rotational velocity $V_C$ 
(measuring the mass of the halo and, by assuming a 
constant halo/disc mass ratio, also the mass of the disc) and spin 
parameter $\lambda$ (measuring the specific angular momentum of the halo).
In fact, a third parameter, the redshift of galaxy formation (depending
on galaxy's mass) is playing a key role in all hierarchical models of galaxy
formation. However, since the term ``time of galaxy formation'' is not
well defined (is it the time that the first generation of stars form? or
the time that some fraction of the stars, e.g. 50\% form?) we prefer
to ignore it and assume that 
{\it all discs  start forming their stars at the
same time, but not at the same rate}. In that case
the profile of a given disc can  be expressed in terms of the one of our
Galaxy (the parameters of which are designated hereafter by index G):
\begin{equation}
\frac{R_d}{R_{dG}}  \  = \  \frac{\lambda}{\lambda_G} \ \frac{V_C}{V_{CG}}
\end{equation}
and
\begin{equation}
\frac{\Sigma_0}{\Sigma_{0G}}  \  =  \left(\frac{\lambda}{\lambda_G}\right)^{-2}
 \ \frac{V_C}{V_{CG}}
\end{equation}
where we have adopted $\lambda_G$=0.03 (see BP99b). Numerical simulations
show that the $\lambda$ distribution peaks around the value 0.04
(e.g. MMW98). The absolute value
of $\lambda_G$ is of little importance as far as it is close
to that peak, since our results depend only  on the ratio $\lambda/\lambda_G$.

Eqs. 5 and 6  allow  to describe the mass profile of a galactic disc
in terms of the one of our Galaxy and of two parameters: $V_C$ and $\lambda$.
The range of observed values for the former parameter
is 80-360 km/s, whereas for the latter
numerical simulations give values in the 0.01-0.1 range, the distribution
peaking around $\lambda\sim0.04$ (MMW98).  
Although it is not clear yet whether
$V_C$ and $\lambda$ are independent quantities, we treat them here as such
and construct a grid of 25 models caracterised by $V_C$ = 80, 150, 220, 290, 360 km/s
and $\lambda$ = 0.01, 0.03, 0.05, 0.07, 0.09, respectively.
[{\it Notice:}  if $\lambda_G$=0.06 is adopted for the Milky Way, our 
model results would be the same, but they would correspond to  values of $\lambda$
twice as large, i.e. 0.02, 0.06, 0.10, 0.14 and 0.18, respectively]. 
Increasing  values of $V_C$ correspond to more massive discs and
increasing values of $\lambda$ to more extended discs.

As discussed in BP99b, the resulting disc radii and central
surface brightness are in excellent agreement with observations, 
except for the case of $\lambda$=0.01.
This ``unphysical'' value leads to  galaxies ressembling
to bulges or ellipticals, rather than discs.
We prefer, however, to keep this
$\lambda$ value in our grid, for illustration purposes.

The two main ingredients of
the model, namely the Star Formation Rate $\Psi(R)$ and the infall time-scale
$\tau(R)$, are affected by the adopted scaling of disc properties
in the following way:

For the SFR we adopt the prescription of Eq. (1), with the same efficiency 
$\alpha$ as in the case of the Milky Way (i.e. the SFR is not a free parameter
of the model).
In order to have an accurate evaluation of
$V(R)$ across the disc, we calculate it as the sum of the contributions 
of the disc (with the surface density profile of Eq.  4) and of the dark 
halo, with a volume density profile of a non-singular isothermal sphere
(see BP99b).

The infall time scale is assumed to increase with both surface density (i.e.
the denser inner zones are formed more rapidly) and with galaxy's mass,
i.e. $\tau[M_d,\Sigma(R)]$.  In both
cases  it is the larger gravitational potential that induces a more 
rapid infall.
The  radial dependence of $\tau$ on $\Sigma(R)$ is calibrated on the Milky Way,
while the mass dependence  of $\tau$ is adjusted as to reproduce
several of the properties of the galactic discs 
(see BP99b).
The adopted prescription allows to keep some simple scaling relations for the
infall in our models and the number of free parameters as
small as possible. The fact that the adopted  simple
prescription provides a satisfactory agreement with several observed 
relationships in spirals (see BP99b) makes us feel that it could be 
ultimately justified by theory or numerical simulations of disc formation.

We stress at this point that in our models we
ignore the possibility of radial inflows in gaseous discs, 
resulting e.g. by viscosity  or by
infalling gas with specific angular momentum different from the one
of the underlying disc; in both cases, the resulting redistribution of angular
momentum leads to radial mass flows. The magnitude of the effect is
difficult to evaluate, because of our poor understanding of viscosity 
and our ignorance of the kinematics of the infalling gas.
Models with radial inflows have been explored in the past
(Mayor and Vigroux 1981; Lacey and Fall 1986; Clarke 1989; 
Chamcham and Tayler 1994); at the present
stage of our knowledge, introduction of radial inflows in the models
would imply even more free parameters and make
impossible  the study of a radial variation in the efficiency of the SFR. 
Since it is well known that bars induce radial mixing and reduce the
abundance gradients (e.g. Dutil and Roy 1999), we do not
consider bared galaxies in what follows.

\section {Results and comparison to observations}

\subsection{Model results}

The evolution of the oxygen abundance profiles in our models is presented in Fig. 1.
Profiles are shown at three ages: T=3, 7.5 and 13.5 Gyr, respectively, covering
most of the galaxian history. It should be noticed that the observed oxygen profile
in the Milky Way disc is nicely reproduced by the corresponding model
($V_C$=220 km/s, $\lambda$=0.03), as analysed in detail in BP99a.
The most important features of Fig. 1 are the  following:

1) Because of the inside-out star forming scheme adopted in our work, the abundance
profile is quite steep early on (when stars and metals are formed only in the inner disc)
and flattens with time (as star formation ``migrates'' progressively outwards).

2) The final abundance gradients (expressed in dex/kpc) are, in general, larger
for the lower mass discs. The reason is that, in such small galaxies, 
the $\Sigma_g^{1.5}/R$
factor  in the adopted SFR can create large differences in abundances 
between regions relatively close to each other. Indeed, because of the small
scalelength of these galaxies, the gas surface density varies a lot within a short
distance, and so does the corresponding SFR. On the contrary, for massive and
large discs, the gas surface density and corresponding SFR vary slowly within one kpc
 and produce flatter abundance gradients (in dex/kpc).

\begin{figure}
\psfig{file=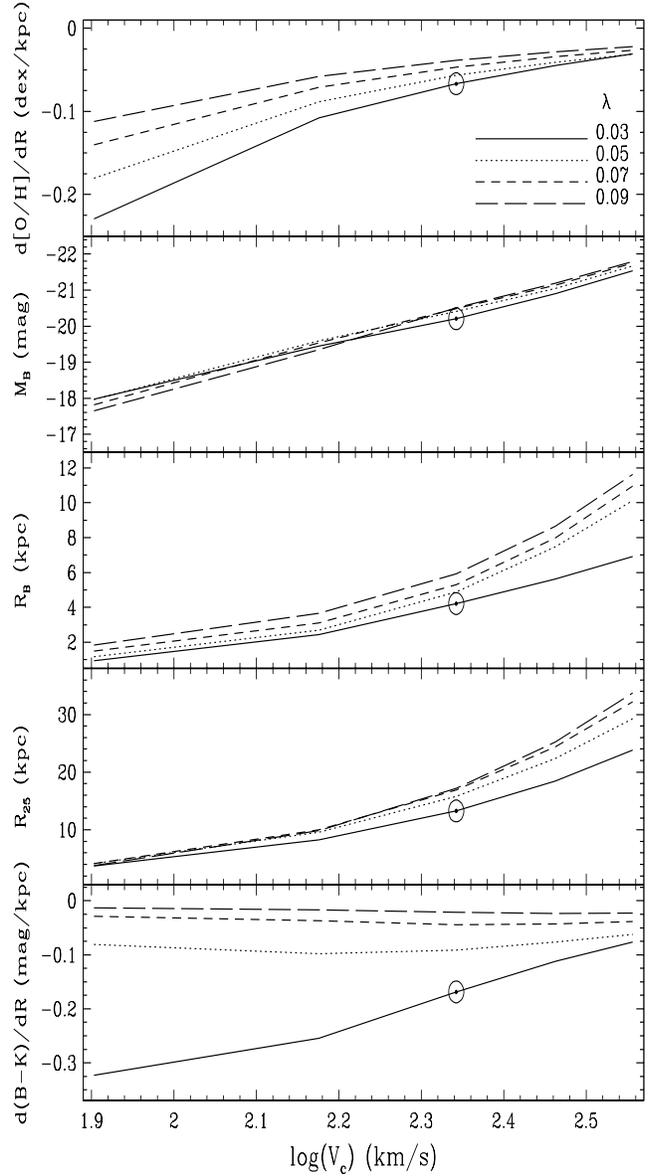,height=17.cm,width=0.5\textwidth}
\caption{\label{} 
Results of our models at an age T=13.5 Gyr. They are plotted as
a function of maximal rotational velocity $V_C$ and spin parameter
$\lambda$ (0.03 {\it solid} curves, 0.05 {\it dotted} curves, 0.07 {\it short-dashed}
curves and 0.09 {\it long-dashed} curves). From  top to bottom: oxygen abundance
gradient (in dex/kpc); absolute B-magnitude M$_B$; disc scalelength in the B-band R$_B$; 
radius where B-band surface brightness is 25 mag arcsec$^{-2}$ (R$_{25}$);
colour gradient B-K (in mag/kpc). In all
panels, the symbol $\odot$ indicates values corresponding to the Milky Way model.}

\end{figure}

3) The more massive the galaxy, the higher is the central (and average) abundance.
This is due to the $V/R$ factor of the adopted SFR, making massive discs more
efficient in forming stars and metals at a given galactocentric distance.
This factor explains also why the most compact discs ($\lambda$=0.01) have very
high abundances. However, as analysed in BP99b, that case does not produce
``physical'' results, since the obtained galaxies ressemble more to ellipticals 
or bulges than to spirals. We show the results for $\lambda$=0.01 in Fig. 1 
for illustration purposes, but in the following we shall discuss only results
for the other cases. In BP99b we show that this increase of metallicity with disc
mass and luminosity compares well to the data of Zaritsky et al. (1994).

\begin{figure*}
\psfig{file=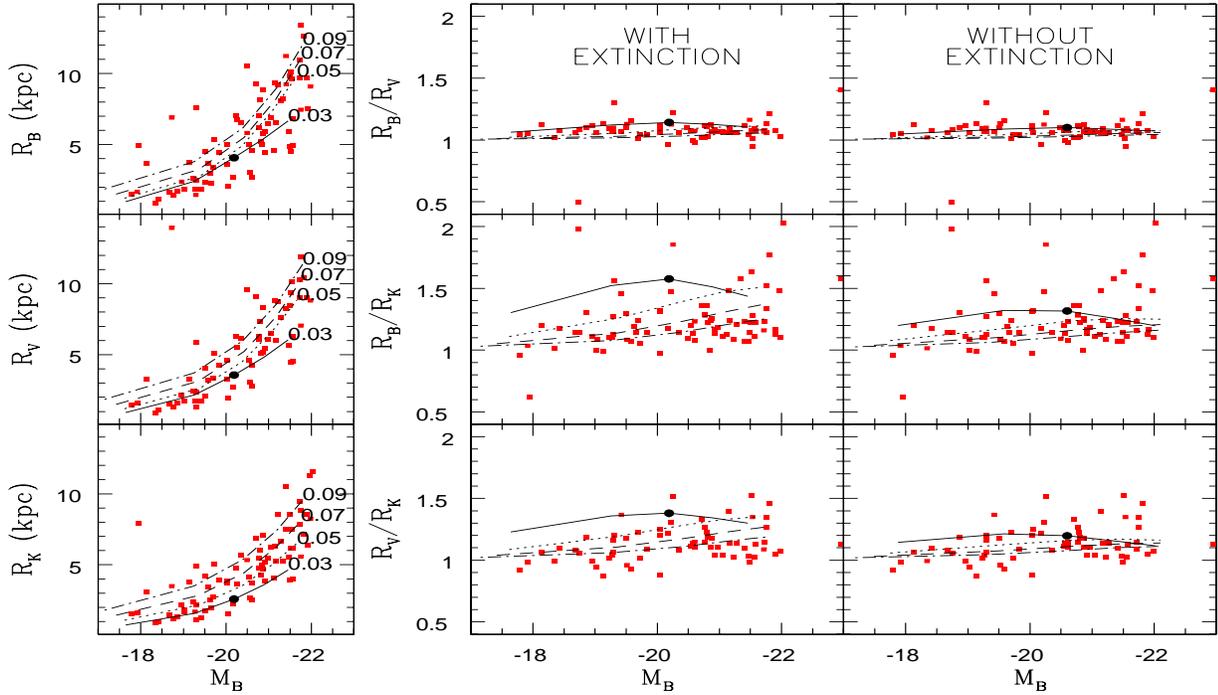,height=10.cm,width=\textwidth,angle=-90}
\caption{\label {}
Disc scalelengths in various wavelength bands ({\it left}) and 
disc scalelength ratios ({\it middle} and {\it right}) as a function of B-magnitude.
Our results ({\it curves}, parametrised with values of the
spin parameter $\lambda$ as indicated on the left) are compared to the
observations of de Jong  (1996). 
Model scalelengths in the {\it middle} panels are calculated with extinction
taken into account, while in the right panels extinction is neglected.
In all panels, the {\it filled circle} 
indicates results for the Milky Way disc.}
\end{figure*}

4) The most efficient galaxies in forming stars and metals are the compact (low $\lambda$) 
and massive (large $V_C$) ones. In their innermost regions, metallicity tends to
saturate, to a value of log(O/H)+12$\sim$9.4. This is a well known 
feature of galactic chemical evolution models taking into account the finite
lifetime of stars
(see Prantzos and Aubert 1995). 
At late times, when most of the gas is exhausted and little 
star formation or metal production takes place, large amounts of metal-poor gas 
are returned to the interstellar medium; they are ejected by the numerous low-mass 
and long-lived stars that were abundantly formed in the early days of the galaxy.
This metal-poor gas dilutes the metal abundances and leads to a ``saturation''
of their value. This feature cannot be revealed by models using the Instantaneous
Recycling Approximation. For that reason, the
oxygen profiles of the $\lambda$=0.01 case are flat, since star formation is very
rapid and efficient all over the (small) disc.

Figure 2 summarizes a number of important features of our models at an age of
T=13.5 Gyr.

a) The oxygen abundance gradient (in dex/kpc) is a monotonically increasing 
function of $V_C$. Notice that gradients are negative and more massive discs
have smaller absolute values of d(log(O/H))/dR, i.e. {\it flatter} abundance
profiles. Also, abundance gradients depend on the $\lambda$ value, especially
in low mass discs: more compact discs have more important (in absolute value)
gradients, while extended discs have flat abundance profiles. In both cases, 
the fact that smaller discs have larger abundance gradients (in dex/kpc)
is due to the $\Sigma^{1.5}$/R factor in the adopted star formation rate,
as explained in point 2 above.

b) The B-magnitude depends mainly on the mass of the galaxy (i.e. rotational
velocity $V_C$) and very little on the spin parameter $\lambda$. As analysed in
detail in BP99b, a Tully-Fisher relation with a slope similar to the observed
one in field spirals and small dispersion is obtained in our models. This is a 
success of semi-analytical models, since the TF relation is ``built-in'' as a
boundary condition (the disc mass being proportional to $V_C^3$).

c) The final B-band scalelength is a monotonically increasing function of
$V_C$ and $\lambda$. This again is due to the ``boundary conditions'' imposed
by the semi-analytic models adopted in our work (see Sec. 2).

% d) Less massive and more compact discs produce not only larger oxygen
% gradients (point a) but also larger surface brightness gradients. This
% fact has interesting observational consequences, as we shall see in Sec. 3.3.

d) R$_{25}$, the radius where B-band surface brightness is 25 mag/arcsec$^2$,
is also found to increase monotonically with $V_C$ and $\lambda$.

e) Finally, discs with $\lambda>0.05$ have small colour gradients, independently
of their mass
(lower panel in Fig. 2). This is because in those galaxies the whole disc
is formed relatively late (in timescales $>$5 Gyr for $V_C <$220 km/s, see Fig.
5 in BP99b) and stellar populations in the inner disc are not much older 
than in the outer regions.

\subsection {Comparison to observations of luminosity profiles}

The exponential scalelength of the Milky Way disc was a subject of
controversy for several years. Indeed, observations in different wavelength bands
lead to different results: a value of R$_B\sim$4-5 kpc is found in the B-band,
while smaller values are found in longer wavelengths (R$_I\sim$2.5 kpc).
In BP99a we showed that these different values result naturally from schemes
of ``inside-out''  formation of the disc, that manage to reproduce the observed 
oxygen abundance gradient of dlog(O/H)/dR$\sim$-0.07 dex/kpc.
In fact, light in the I- or K- bands traces closely the old stellar population,
which dominates the mass of the disc in all radii. Light in the B-band
traces a younger stellar population, formed in the past 1-2 Gyr from a gas with
a surface density profile flatter than the one of the old population.
Thus, colour and abundance gradients are found to be naturally correlated
in the case of the Milky Way disc.

In this section we explore the results of our models concerning the disc scalelengths
in various wavelength bands. Our aim is: 1) to see whether our results are
compatible with the observations and 2) to check whether there are any similarities
with the case of the Milky Way.

On the left part of Fig. 3 we present the disc scalelength of our models 
in the B-, V- and K- bands (evaluated with the effects of
extinction taken into account)
and compare them to the observations of De Jong (1996).
De Jong  has decomposed the luminosity profiles of his  $\sim$80 face-on galaxies
in bulge and disc components and we plot in Fig. 3 only the corresponding disc
component, directly comparable to our results. It can be seen that the model
disc scalelengths are in excellent agreement with the observed ones as a function
of M$_B$. More massive and bright discs have larger scalelengths on the average, while
the variation of the spin parameter $\lambda$ allows to reproduce
the observed scatter (perhaps not quite efficiently in the B-band).

On the right part of Fig. 3 we plot the disc scalelength ratios as a function
of M$_B$. Since extinction may affect the results, we plot our results with
extinction included (middle panels) and neglected (right panels).
It can be seen that the observed R$_B$/R$_K$ and R$_V$/R$_K$ rates are on the
average larger than 1 (i.e. there is a colour gradient); also, there is
a marginal trend of  increasing ratios with galaxy luminosity. Our results
are in agreement with the
observed trends, in particular when extinction is included.
We find that the case of $\lambda$=0.03 leads to the most extreme scalelength
ratios; other $\lambda$ values (i.e. less compact discs) produce more modest
colour gradients. In that respect then, {\it our Milky Way is a rather compact disc
having larger than average colour gradients}.

\subsection {Comparison to observations of abundance gradients}

In this section we compare our results to the observations of Garnett et al. (1997)
and van Zee et al. (1998) concerning abundance gradients in external spirals 
(i.e. observations of O/H in HII regions as a function of galactocentric distance). 

\begin{figure}
\psfig{file=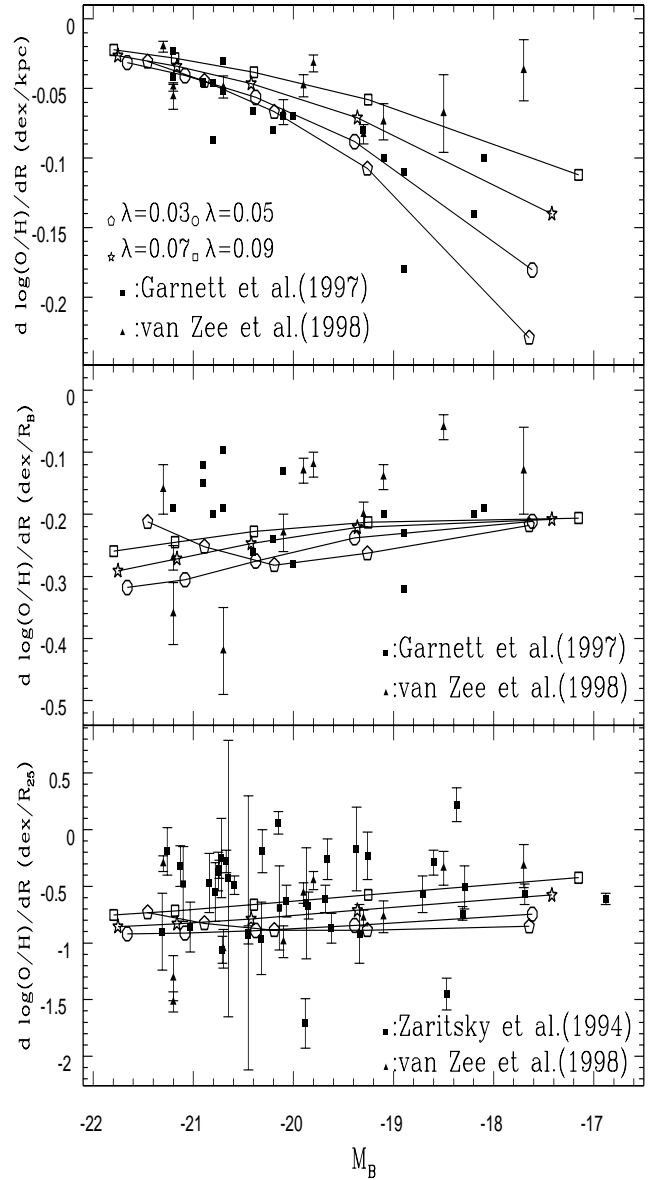,height=17.cm,width=0.5\textwidth}
\caption{\label {}
Oxygen abundance gradients of  our models at T=13.5 Gyr ({\it open symbols}, corresponding
to the values of the spin parameter $\lambda$ as indicated in the upper
panel and connected with {\it solid} curves). They are plotted as a function
of the B-magnitude and compared to observations ({\it filled symbols}, with 
appropriate references given in each panel).
{\it Upper panel}: Oxygen gradient in dex/kpc;
{\it Middle panel}: Oxygen gradient in dex/R$_B$;
{\it Lower panel}: Oxygen gradient in dex/R$_{25}$.}

\end{figure}

Garnett et al. (1997) combine data obtained in several previous works (Vila-Costa
and Edmunds 1992, Zaritsky et al. 1994 and Ryder 1995) and make several adjustements
in order to introduce a greater degree of uniformity. From this final sample we
selected a sub-sample of 17 non-barred galaxies, since those compare better
to our models (bars can induce mixing of gas along the disc and flattening
of abundance gradients in relatively short timescales, e.g. Friedli et al. 1998
and references therein; our models do not take such effects into account).
On the other hand, van Zee et al. (1998) combine data from the literature
(including their own observations for low metallicity HII regions) and derive
abundance gradients for 11 more spirals.

Our results are compared to observations in Fig. 4, where abundance gradients are
plotted as a function of integrated B-magnitude.

In the upper panel, abundance gradients are expressed in dex/kpc. Observations show
that luminous discs have small gradients, while as we go to low luminosity ones
the absolute value of the gradients and their dispersion increase.
This trend is fairly well reproduced by our models. The reason can be understood 
by comparing to the upper panel of Fig. 2, where our results are plotted as
a function of $V_C$ (which correlates well to M$_B$ through the Tully-Fisher
relation, BP99b). As analysed in Sec. 3.1, the $\Sigma_g^{1.5}$/R factor in
the adopted SFR creates large abundance variations within short distances in
small galaxies; no important gradients can be created in massive discs, 
where neighboring 
regions differ little in $\Sigma_g$. As for the increased dispersion at low
luminosities (and low $V_C$), it is due to the effect of the $\lambda$ parameter:
for large $\lambda$ values, even small mass discs are quite extended
(Eq. 5 in Sec. 2) and the $\Sigma^{1.5}$/R factor becomes less and less
efficient in creating abundance gradients (since $\Sigma$ varies little as a function of R). Thus, {\it in low mass discs, both the
mass and the spin parameter $\lambda$ drive the abundance gradient.}

As stressed in BP99b, discs with $\lambda>0.08$ correspond to Low Surface 
Brightness galaxies (LSBs, see also Jimenez et al. 1998). According to 
DeBlok and van der Hulst (1998) LSBs have negligible abundance gradients. Our results
of Fig. 4 are quite encouraging in that respect: indeed, had we run models with
$\lambda>0.09$ we would expect to find smaller and smaller abundance gradients
at all galaxy luminosities.
We postpone, however, the detailed study of LSBs in a forthcoming paper.

In the middle panel of Fig. 4, the oxygen gradients are expressed in dex/R$_B$.
When expressed in this unit, the observed abundance gradients show no more any 
correlation to M$_B$, as already noticed in Garnett et al. (1997). Moreover,
a considerable dispersion is obtained for all M$_B$ values,
while the average gradient is $\sim$-0.2 
dex/R$_B$. Our results (values in the upper panel of Fig. 2, multiplied by
the corresponding R$_B$ values of the middle panel of Fig. 2) show also
no correlation with M$_B$, in fair agreement with observations.
However, the average value is slightly larger than observed 
($\sim$-0.25 dex/R$_B$).
Since the estimates of R$_B$, both in observations and in our models,
may be affected by extinction, we consider that the agreement with the data
is quite satisfactory. Another difficulty may stem from the fact that
relatively few discs can be fit with perfect exponentials; according to
Courteau and Rix (1999) this happens for only $\sim$20\% of the disc galaxies.
Finally, we obtain no significant dispersion in our models, contrary to 
obervations. We suspect that the difficulties to evaluate the 
effects of extinction or to
fit the profiles with a perfect exponential may be (at least partly)
responsible for the  observed scatter.

In the lower panel of Fig. 4 abundance gradients are expressed in 
dex/R$_{25}$. Again, observations show no trend with M$_B$ and a
considerable dispersion. 
Our results also show no correlation with M$_B$,
while the average model value ($\sim$-0.8 dex/R$_{25}$) is in perfect
agreement with observations. 
This is most encouraging, since $R_{25}$ is less affected by 
considerations on dust extinction or fit to an exponential
profile. Indeed, we think that the obtained average value of 
dlog(O/H)/dR=-0.8 dex/R$_{25}$ is a more robust result than the 
-0.25 dex/R$_B$ value obtained in the previous paragraph. However, the
observed dispersion is not reproduced by our models; only part of it can
be attributed to the effects of the spin parameter $\lambda$.

\begin{figure}
\psfig{file=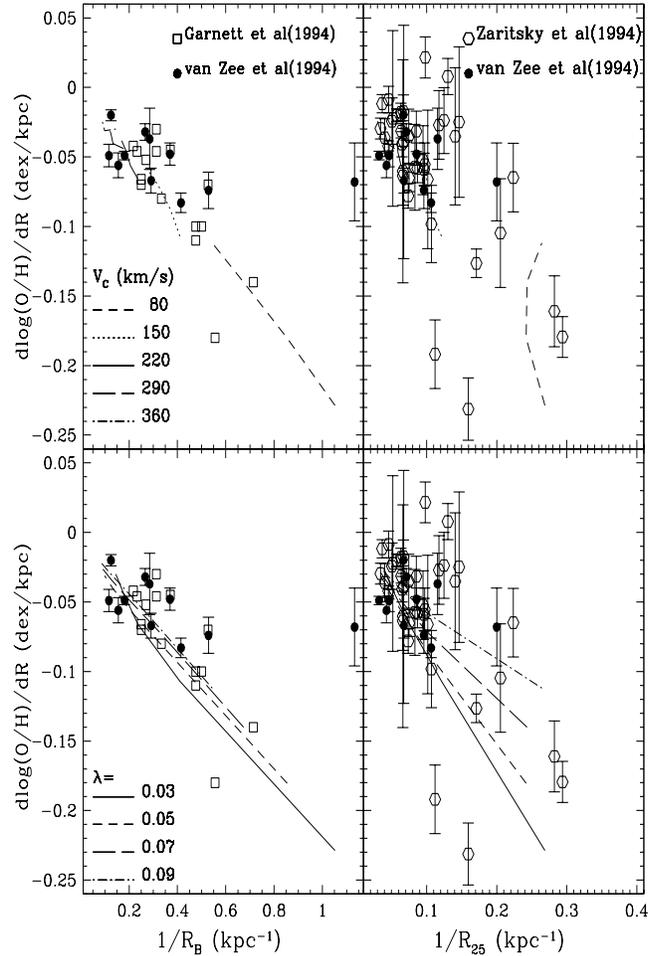,height=14.cm,width=0.5\textwidth}
\caption{\label{}
Predictions of our models for the gradients of metallicity (in dex/kpc) vs 
1/R$_B$ (left panels) and 1/R$_{25}$ (right panels). In the upper panels
our models are parametrised by $V_C$ values and in the lower panels by
$\lambda$ values. Data are from Garnett et al. (1997), van Zee et al. (1998)
and Zaritsky et al. (1994). A clear correlation is found between abundance 
gradient (in dex/kpc) and 1/R$_B$.}

\end{figure}

\subsection{Abundance and colour gradients vs. scalelength}

In the previous section it was shown that:

a) abundance gradients (in dex/kpc) become more important and present
a larger dispersion in low luminosity discs.

b) Abundance gradients in dex/R$_B$ are independent of disc luminosity, and so
is the corresponding dispersion.

Points (a) and (b) immediately suggest that there must be a one-to-one
correlation between abundance gradients and scalelength R$_B$.
The correlation must be relatively tight, since the 
observed dispersion (middle panel of Fig.  4) is relatively small.
Although this conclusion is an immediate consequence
of the observations in Fig. 4, it has never been stated explicitly 
(to our knowledge). For the same reason, a one-to-one correlation must 
exist between abundance  gradients and R$_{25}$. 
However, the observed scatter (lower panel in Fig. 4) shows that 
one should not expect a very tight correlation in that case.

In Fig. 5 we plot the oxygen abundance gradients (in dex/kpc) as a function of 
1/R$_B$ 
(which is proportional to a ``magnitude gradient'' in mag/kpc) 
and of 1/R$_{25}$, both from observations and from our  models. 
In the upper panels our results are parametrised by the values of rotational
velocity $V_C$, while in the lower panels they are parametrised by the 
values of the spin parameter $\lambda$. It can be seen that
our models show an excellent correlation between the abundance gradients and
1/R$_B$: smaller discs have larger gradients and the results depend
little on $\lambda$. In principle, knowing the B-band scalelength of a disc,
one should be able to determine the corresponding abundance gradient and 
vice-versa! In practice, however, observations show a scatter around the
observed trend. Moreover, our results seem to agree better with the data
of Garnett et al. (1997) than with those of van Zee et al. (1998). The
latter show a less steep dependence of the gradient on 1/R$_B$. Since large
discs ``cluster'' on the upper left corner of the diagram, we
suggest that observations of the abundance gradients in small discs would
help to establish the exact form of the correlation.
Finally, we notice that in our models we find similar correlations
between abundance gradients and disc scalelengths in other 
wavelength bands (as can be expected from the fact that scalelength ratios
are close to unity, Fig. 3).

The situation is not as promising when abundance gradients are plotted as 
a function of 1/R$_{25}$. Our models show again a correlation, but this time the
different $\lambda$ values lead to increasingly different abundance gradients
as one goes to smaller discs. Although the data are compatible with the
obtained relation, the large observational 
uncertainties do not allow to conclude. Again,
further observations for small discs are required to clarify the situation.

\begin{figure}
\psfig{file=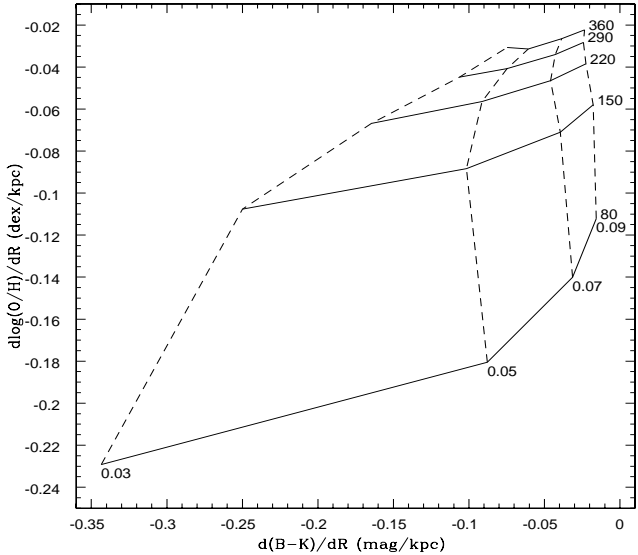,height=8.cm,width=0.5\textwidth}
\caption{\label{} 
Predictions of our models at T=13.5 Gyr for the relation between metallicity
gradient (in dex/kpc) and colour gradient B-K (in mag/kpc). The grid of our
models is parametrised by $V_C$ (360, 290, 220, 150 and 80 km/s, from 
top to bottom) and $\lambda$ (0.03, 0.05, 0.07 and 0.09, from left to right).
Massive and extended discs have negligible metallicity and colour gradients.}

\end{figure}

Finally, our models predict that, in general, colour gradients are developped
in discs, because of the inside-out star formation scheme. For instance, in
the case of the Milky Way we obtain R$_B\sim$4 kpc and R$_K\sim$2.6 kpc,
in agreement with observations (see BP99a and references therein).
As shown in Fig. 2 (bottom panel) and discussed in Sec. 3.1, colour gradients are 
rather small for discs with $\lambda>$0.05; only for low $\lambda$ and 
$V_C$ values do they become important. In Fig. 6 we present the obtained
relationship between abundance and B-K gradient, expressed in dex/kpc 
and mag/kpc, respectively. Although the resulting grid covers a rather large part
of the figure, it nevertheless excludes the co-existence of large colour
and small abundance gradients. Combined chemical and photometric analysis
of disc profiles should allow to find out whether the proposed scheme is realistic
or not.

\section{ Summary}
 
In this work, we study the photometric and chemical abundance profiles resulting
from our models of the evolution of galactic discs. Our 
galaxy evolution models are calibrated on the Milky Way (BP99a)
and reproduce a  large body of observational data concerning integrated
properties of external spirals (BP99b).
Here it is shown that they also produce results in excellent agreement with 
observations of abundance gradients in spiral galaxies. More specifically:

1) Because of the inside-out star formation scheme, our models predict that
abundance and photometric profiles flatten with time.

2) Model abundance gradients (in dex/kpc) are more important and present a larger
dispersion for low luminosity discs, in fair agreement with observations.
We claim  that the spin parameter $\lambda$ (affecting the size, but not the
mass of the disc in our models) is responsible for the scatter. Discs with larger
$\lambda$ values than those studied here should appear as Low Surface Brightness
galaxies and should exhibit even smaller abundance 
gradients than those displayed in Fig. 3.
Observations  show indeed that LSBs have negligible abundance gradients 
(DeBlok and van der Hulst, 1998). 

3) When expressed in dex/R$_B$ and dex/R$_{25}$, observed abundance gradients show 
no dependence on galaxy luminosity; they do show an important scatter, especially
in the latter case. Our models also show no trend with luminosity, while the
average values are in fair agreement with observations. We fail, however, to
reproduce the observed scatter (variations induced by the different $\lambda$ values
are not sufficient for that).

4) The observed and theoretical relation between galaxy luminosity and abundance 
gradients, suggest that there must also exist a correlation between the latter and
1/R$_B$ or 1/R$_{25}$. Our models show indeed such a correlation: abundance gradients
(in dex/kpc) correlate much better to 1/R$_B$ than to M$_B$. Observations seem to
coroborate this idea, although more data are needed to establish the precise form
of that correlation.

5) Our models show no correlation between abundance and colour gradients (expressed
in dex/kpc and mag/kpc, respectively), except for the most compact discs
(those with $\lambda$=0.03). Compact discs of low luminosity are found to have
large abundance and colour gradients.

The most interesting feature of this work is the anticorrelation found between abundance
gradients and scalelength.{ \it Discs with small scalelengths have large abundance gradients,
independently of their mass.} 
It is the first time (to our  knowledge) that  this relationship is noticed.
It should be interesting to establish it through further observations,
concerning in particular discs of small scalelengths.
Our models seem to reproduce that relationship quite naturally (Fig. 5).

We wish to stress here that our simple chemical evolution models treat galactic discs
as ensembles of concentric, independently evolving rings, i.e. no radial
flows ar considered. If radial inflows are indeed playing a key role in shaping
abundance gradients (as sometimes claimed), how is then the success of our 
models to be understood?

As explained in Sec. 2, our models are ``calibrated'' 
on the Milky way disc. In particular, the radial dependence of the adopted infall 
rate is such that, when combined to the adopted SFR, reproduces the current profiles 
of gas, stars and metal abundances.  The fact that we keep the same prescriptions
for the SFR and infall rate in the other disc models, leads obviously to a kind
of ``homologuous evolution'', as already suggested in Garnett et al. (1997).
In other words, prescriptions adopted for the Milky Way are found to be equally
successful when applied to other discs.
It should be interesting to see whether hydrodynamical
simulations (like e.g. those of Samland et al. 1997, or Contardo et al. 1998)
manage to produce this kind of ``homologuous evolution'' in galactic 
discs. We notice, however, that the idea of a ``homologuous evolution'' 
merely states a fact, but does not itself explain the origin of the observed 
gradients.

\label{lastpage}

\end{document}